\documentstyle[prl,aps,epsf,floats,twocolumn]{revtex}

\begin{document}
 

\wideabs{

\title{ Non-volatile magnetoresistive memory in phase separated La$_{0.325}$Pr$_{0.300}$Ca$_{0.375}$MnO$_3$.}
\author{ P. Levy,$^{+,*}$ F.Parisi,$^{b}$ M.Quintero, L.Granja, J.Curiale, J.Sacanell,
G.Leyva,$^{b}$ G.Polla}
 
\address{Departamento de F\'\i sica, Comisi\'{o}n Nacional de
Energ\'\i a At\'{o}mica,Gral Paz 1499 (1650) San Mart\'\i n,Buenos
Aires,Argentina}

\author{R.S. Freitas and L.Ghivelder.}

\address{Instituto de F\'\i sica, Universidade
Federal do Rio de Janeiro, C.P. 68528, Rio de Janeiro, RJ 21945-970, Brazil }
\maketitle

\begin{abstract}

We have measured magnetic and transport response on the polycrystalline La$_{5/8-y}$Pr$_y$Ca$_{3/8}$MnO$_3$ ($y=0.30$,
average grain size 2 microns ) compound. In the temperature range where ferromagnetic metallic and insulating regions coexist
we observed a persistent memory of low magnetic fields ( $<$ 1 T) which is determined by the actual amount of the ferromagnetic
phase. The possibility to manipulate this fraction with relatively small external perturbations is related to the phase
separated nature of these manganese oxide based compounds. The colossal magnetoresistance figures obtained (about 80\%) are
determined by the fraction enlargement mechanism. Self-shielding of the memory to external fields is found under certain
described circumstances. We show that this non-volatile memory has multilevel capability associated with different applied
low magnetic field values. 

\end{abstract}


Acepted for publication in Phys. Rev. B: Rapid Comm.

$^+$ corresponding author levy@cnea.gov.ar}

\narrowtext

The unusually large change of resistivity following application of magnetic field, the so called colossal magnetoresistance
(MR) effect  observed in manganese-oxide-based compounds, is again being the focus of intense research due to the capability
of small external forces to tailor the macroscopic material´s response. This unique possibility is related to the
simultaneous presence of submicrometer ferromagnetic metallic (FM) regions and charge ordered (CO) and/or paramagnetic
insulating (PI)ones in some manganites, the Phase Separation (PS) phenomena.\cite{Dagotto}

The PS scenario fully develops in the  La$_{5/8-y}$Pr$_y$Ca$_{3/8}$MnO$_3$ (LPCM($y$)) family of compounds, whose end members
exhibit homogeneous FM ($y$=0) and CO ($y$=1) states. \cite{Khomskii} Intense research activities in these compounds were
developed by Moscow groups \cite{Babushkina,Balagurov,Voloshin} and by Cheong\'s group. \cite{Uehara,UeharaRate,Kim,Podzorov1/fnoise,PodzorovFluctu,PodzorovMarten,Kiryukhin}
In short, these works show the tendency of LPCM($y$) to form inhomogeneous structures, which seems to be a generic property
of strongly correlated systems. \cite{Khomskii,Mathur}

There are conclusive evidences, both theoretical \cite{MoreoPRL,Georgios}and experimental, \cite{Parisi,Luis} showing that
the high values of  MR achieved by PS manganites are due to the possibility of unbalancing the amount of the coexisting
phases by the application of low magnetic fields. In La$_{0.5}$Ca$_{0.5}$MnO$_{3}$, a prototypical PS compound with its
charge ordering temperature $T_{co}$ lower than that of FM ordering $T_{C}$, the huge values of the low field low temperature
MR are related to the capability of the field to inhibit the formation of otherwise CO regions, leading to an effective
increase of the FM fraction. \cite{Parisi} Remarkably, it  was also shown that  this process is only achieved when field
cooling (FC) the sample. \cite{Parisi} Application of low $H$ once the relative fractions of the coexisting phases were
established has the only effect of domain alignment. At low temperature, this kind of MR affects mainly the
intergrain tunneling in polycrystalline samples, with maximum values of the low field MR around 30\%.
\cite{LeeIntergrain,HwangSPIT} The application of moderate magnetic fields in the FC mode plays also an unbalancing role in
Cr-doped Nd$_{0.5}$Sr$_{0.5}$MnO$_{3}$\cite{Kimura}: the low temperature zero-field resistivity of this compound display a
strong dependence with the cooling field. The reduction in the resistivity persists even after this annealing field is
removed, meaning  that the magnetic field remains imprinted in the sample\'s resistivity.

The possibility of producing technological applications with PS manganites, both for magnetic reading heads and
magnetoresistive
data storage,  has been one of the aims in the investigation of these kind of systems. However, the mentioned features
related to the way in which colossal values of MR can be achieved or the magnetic fields can be imprinted, seem to impose
drastic restrictions to its actual implementation.

In this work we show that  low magnetic fields can be effectively written under isothermal conditions in the PS manganite
 La$_{0.325}$Pr$_{0.300}$Ca$_{0.375}$MnO$_3$. This effect is achieved due to the confluence of two factors. On one hand, the
fact that within a
delimited temperature range the equilibrium state of the system is a true PS one with definite equilibrium fractions of the
coexisting phases. On the other one, the fact that in this temperature range the system displays out of equilibrium features
governed by a slow dynamics, allowing different long term metastable states to be tuned by a small external stimulus. We
claim that the
identification and understanding of these ingredients produces a breakthrough  for technological uses of  PS manganites.

Polycrystalline samples of LPCM(0.3) were synthesized by the sol - gel technique. Thermal
treatments were performed at 1000 C. Average grain size determined through Scanning Electrom Micrographs was around 2
microns. Resistivity $\rho$ (standard four probe technique) and magnetization $M$ (extraction QD PPMS magnetometer) were
measured as a function of temperature T in the presence of an external low magnetic field ($LH$) $H$ $<$ 1 T.

Figure 1 displays magnetization, resistivity and magnetoresistance as a function of temperature for a LPCM(0.3) sample.
Around 220 K the CO state develops, as evidenced both by the increase in $\rho$ and the peak in $M$. Some degrees below, at
$T_{C1}\approx $ 210 K, FM clusters nucleate within the CO matrix,  producing a peak in $\rho$($T$) and a steep increase in
$M$($T$). The FM volume fraction of this state, which extends down to $T_{C2}\approx$ 100 K, is nearly constant (about 14 \%)
i.e., below the percolation threshold (17\%).\cite{Kim} In this temperature range the state of the system is characterized by
the coexistence of frozen isolated  FM clusters within CO regions; these clusters start to grow at $T_{C2}$, and an insulating
to metal transition develops when the fraction of the FM phase reaches the percolation threshold. Below the
freezing temperature $T_{f}\approx$ 75 K the $M$($T$) data exhibits a plateau, revealing that no further evolution
in the fraction of the coexisting phases is produced on further cooling. The FM volume fraction in this low $T$ region is
about 90 \%. These results are in good agreement with previously reported data on the LPCM(0.3) compound. \cite{Uehara,Kim}
\begin{figure}[t]
\centering
\epsfysize=10.0 cm
\epsffile{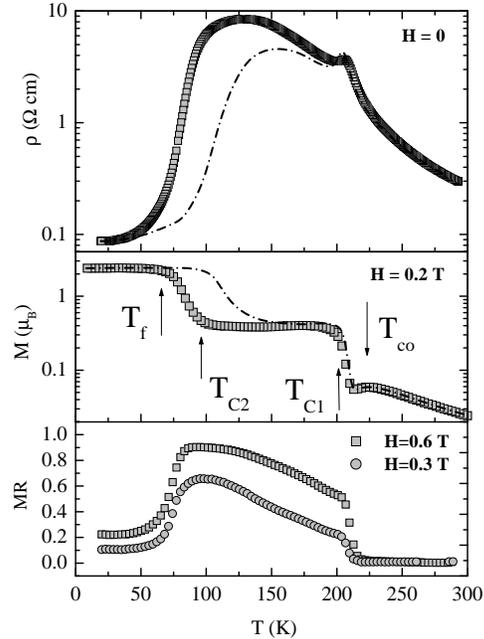}
\caption{Temperature dependence of a) $\rho$($H$ = 0), b) $M$ ($H$ = 0.2 T) and c) MR = ($\rho$ (0) - $\rho$($H$))
/
$\rho$(0) on cooling (filled symbols) and warming (dashed line) for polycrystalline LPCM(0.3); $H$ values for the MR data are shown in the graph. 
\label{Fig1}
}
\end{figure}
The MR = $(\rho(0) - \rho(H)) / \rho(0)$ obtained on cooling (FC procedure) is depicted in Fig. 1c for different $LH$ values.
The high ( $>$ 60\%) MR values obtained within the PS regime occurring between $T_{C1}$ and $T_{C2}$ can be understood as the
result of the enlargement of the FM phase induced by the LH. \cite{Parisi} The increase of the FM fraction occurring on
cooling down through $T_{C2}$ is concomitant with the decrease of the MR curve. This fact can be also understood within the
fraction enlargement mechanism: low MR values  are obtained as soon as  the FM fraction departs from the percolation
threshold. \cite{Parisi} Below $T_{f}$ the fraction of the coexisting phases is no longer controlled by the external field:
the low temperature MR figures (~ 30 \% at $H$ = 0.6 T)  are consistent with those expected for an homogeneous
polycrystalline system. \cite{HwangSPIT} So,  no changes in the relative fraction of the coexisting phases are induced by
$LH$ below the freezing temperature.

The unusual inhomogeneous state that develops in the $T$ range between $T_{C1}$ and $T_{C2}$ was previously shown to be
characterized by interesting time dependent effects, as cooling rate dependence \cite{UeharaRate}, relaxation in transport
\cite{Babushkina} and magnetic \cite{Voloshin} properties, giant 1/f noise, \cite{Podzorov1/fnoise} non equilibrium
fluctuations, \cite{PodzorovFluctu} etc. The microscopic origin of the dynamics observed in manganites with coexisting FM and
CO
regions (either short or long ranged) could be related to the strain accommodation \cite{Mathur}, which is induced as a
result of martensitic nucleation of the child phase within the parent matrix. \cite{PodzorovMarten}

Time relaxation of both $\rho$ and $M$ were observed between $T_{C1}$ and $T_{f}$, indicating that, in this $T$
range, the data shown in Fig 1 does not correspond to the equilibrium state. This fact is in close agreement with previously
reported cooling rate dependences. \cite{UeharaRate} The slope of the relaxations (positive
for M measurements, negative for the $\rho$ ones) signals that the system reaches the temperature T with a FM fraction lower than
that corresponding to the equilibrium state.  In Fig. 2 we shown the effect of $LH$ applied after zero field cooling the
sample to a
temperature close below $T_{C2}$, while the system is relaxing towards equilibrium. Fig. 2a displays $\rho$ at $T$= 95.5 K as
a function of elapsed time upon the application of magnetic fields of 0.1, 0.2, 0.3, 0.4 and 0.5 T with intermediate
switching off.  Sudden decreases are observed in $\rho$ when the field is applied, followed by slow relaxations. These jumps are
related to both the alignment of spins and domains with the field  and to the enlargement of the FM phase
driven by $H$. Remarkably, when the field is turned off the resistivity steeply increases, but without recovering its
previous $\rho_{H=0}$ (i.e. $\rho$ when $H$=0) value. As it is suggested by the staircase structure displayed in Fig. 2a, the
combination of the  mentioned MR effects determines a response which is persistent after the magnetic field is removed.

An interesting fact is that the system still relaxes after the field is switched off (this is not apparent from Fig. 2a due
to the time scale used). The inset of Fig. 2a compares the normalized $\rho_{H=0}$ values extracted from Fig. 2 as a function
of the elapsed time after the $LH$ was turned off. Also remarkably, there exists a threshold value $H_{th}$ above which the
relaxation of $\rho(t)$ reverses its sign, i.e. $\rho (t)$ slowly increases (instead of decreasing) upon increasing the value
of the $LH$ applied. This result suggests that the amount of the FM phase decreases after $H$ $>$ $H_{th}$ is applied, as if
the system was driven to an over - enlarged metaestable state. This fact signs unambiguously  that the equilibrium state of
the system between $T_{C2}$ and $T_f$ is of PS nature, characterized by an equilibrium fraction of the FM phase.
\begin{figure}[t]
\centering
\epsfysize=10.0 cm
\epsffile{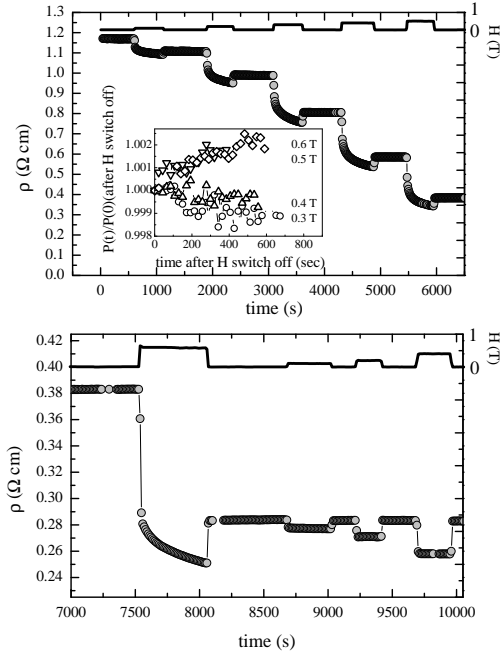}
\caption{ a) $\rho$(95.5 K) as a function of time upon application of $H$ = 0.1, 0.2, 0.3, 0.4 and 0.5 T. Inset: time
dependence
of $\rho$(95.5 K) in H = 0  normalized to the $\rho$ value after the field $H$ was turned off; the labels are the
$H$ fields; b) $\rho$(95.5 K) as a function of time upon application of H = 0.1, 0.2, 0.4 after $H_{MAX}$ = 0.6 T has
been applied and turned off.
\label{Fig2}
}
\end{figure}  

To further elucidate the contribution of each MR mechanism (alignment and enlargement) to the overall persistent MR effect,
we studied the response of $\rho_{H=0}$ (determined by the previous application of some $H_{MAX}$) when further applying $H$
$<$ $H_{MAX}$. Fig. 2b shows the effect of application and removal of  0.1, 0.2 and 0.4 T after $H_{MAX}$ = 0.6 T have been
applied and removed at T = 95.5 K.  As seen, when $H$ is turned on, an immediate response producing a decrease of $\rho$ is
observed, which vanishes when $H$ is switched off. The staircase structure of Fig. 2a is now replaced by deeps at the
intervals when the field is turned on, their depth being determined by the $LH$ strength, with a characteristic specific
MR/$H$ value of 22 \% / T.

By simple inspection of the above presented results the persistent effect on the resistivity can be unambiguously assigned to
the FM enlargement process. The non-trivial relaxation of the system towards its equilibrium point by increasing the FM
volume fraction indicates the existence of a distribution of energy barriers through which the FM phase grows against the
non-FM one. \cite{Granja} As soon as a field $H_{MAX}$ is applied the system is forced to increase its FM fraction overcoming
the barriers
of height less than $\mu H_{MAX}$  ($\mu$ a characteristic parameter of the system relating the field and energy
scales).  Once the system is driven to a "close to equilibrium" state, the effect can not be reversed, and the system keeps
memory of its magnetic history in the FM phase fraction. Within this framework, no additional FM enlargement is produced by
the ulterior application of $H$ $<$ $H_{MAX}$, and the only effect achieved is due to the alignment mechanism. When $H$=0
this alignment MR vanishes, and the response associated with the same previously existing FM volume fraction (determined by
$H_{MAX}$) is restored, i.e. the previous $\rho_{H=0}$ is obtained (Fig. 2b). Thus, the electrical transport response
$\rho_{H=0}$ is not altered by processes involving external $H$ as long as $H$ $<$ $H_{MAX}$; the system keeps memory of
$H_{MAX}$ encoded in its resistivity in a state which is shielded against external H$<$ $H_{MAX}$.

The above described scenario seems to be characteristic of the so called "low $T_C$" manganites exhibiting PS. We obtained
very similar data to that shown in Fig.2 on  La$_{0.5}$Ca$_{0.5}$Mn$_{1-y}$Fe$_{y}$O$_{3}$ for 0.2 $<$ $y$ $<$ 0.6, another PS
family of compounds with rather different hole doping level. \cite{Novel} The key ingredients leading to the magnetoresistive memory are
the existence of a true PS thermodynamic state joined to the possibility of tuning long term metastable states by small
external stimuli. As a distinctive fact, the strength of the persistent MR effect is directly related to the magnitude of the
$LH$ applied, envisaging applications as a sort of "analogical memory device". This multilevel memory associated with
different applied magnetic fields below 1 T has similarities to that achieved using electric fields in thin oxide films with
pervskite-like structure.
\cite{Beck} In our case, the system keeps memory of its magnetic history in the FM phase fraction; it  can be recovered by
transport measurements, with a MR/H performance of  $\approx$ 80 \% / T, opening a path for the design of non-volatile
magnetoresistive memories.

--------------------
Project partially financed by CONICET, Fundaciones Antorchas,  Balseiro and Vitae.

$^b$ Also at ECyT, Universidad Nacional de Gral. San Martín, San Martín, Argentina.
$^*$ Also at CIC, CONICET, Argentina.

\end{document}